\documentclass
[superscriptaddress,secnumarabic,amssymb,amsmath,nobibnotes,aps,prd,nofootinbib,column,notitlepage,10pt]{revtex4}%
\usepackage{setspace}
\usepackage{color}
\usepackage{amsmath}
\usepackage{amsfonts}
\usepackage{verbatim}
\usepackage{amssymb}
\usepackage[colorlinks]{hyperref}
\usepackage{graphicx,bm}
\usepackage{amsmath}
\usepackage{amssymb}
\usepackage{graphicx}
\usepackage{textgreek}
\usepackage[caption=false]{subfig}%
\providecommand{\U}[1]{\protect\rule{.1in}{.1in}}

\newcommand{\be}{\begin{equation}}
\newcommand{\ee}{\end{equation}}

\newcommand{\mincir}{\raise
-3.truept\hbox{\rlap{\hbox{$\sim$}}\raise4.truept\hbox{$<$}\ }}
\newcommand{\magcir}{\raise
-3.truept\hbox{\rlap{\hbox{$\sim$}}\raise4.truept\hbox{$>$}\ }}

\ifx\pdfoutput\relax\let\pdfoutput=\undefined\fi
\newcount\msipdfoutput
\ifx\pdfoutput\undefined\else
\ifcase\pdfoutput\else
\ifx\paperwidth\undefined\else
\ifdim\paperheight=0pt\relax\else\pdfpageheight\paperheight\fi
\ifdim\paperwidth=0pt\relax\else\pdfpagewidth\paperwidth\fi
\fi\fi\fi

\begin{document}
\title{Observational constraints on the modified cosmology inspired by string T-duality}
\author{G.G. Luciano}
\email{giuseppegaetano.luciano@udl.cat}
\affiliation{Departamento de Qu\'{\i}mica, F\'{\i}sica y Ciencias Ambientales y del Suelo,
Escuela Polit\'{e}cnica Superior -- Lleida, Universidad de Lleida, Av. Jaume
II, 69, 25001 Lleida, Spain}
\author{A. Paliathanasis}
\email{anpaliat@phys.uoa.gr}
\affiliation{Department of Mathematics, Faculty of Applied Sciences \& Institute of Systems Science, Durban University of Technology, Durban 4000, South Africa}
\affiliation{National Institute for Theoretical and Computational Sciences (NITheCS), South Africa}
\affiliation{Departamento de Matem\`{a}ticas, Universidad Cat\`{o}lica del Norte, Avda.
Angamos 0610, Casilla 1280 Antofagasta, Chile}
\author{A. Sheykhi}
\email{asheykhi@shirazu.ac.ir}
\affiliation{Department of Physics, College of Science, Shiraz University, Shiraz 71454, Iran}
\affiliation{Biruni Observatory, College of Science, Shiraz University, Shiraz 71454, Iran}
\begin{abstract}
We explore the cosmological consequences of a modified cosmology
inspired by string T-duality. We incorporate the zero-point length
correction, $l_0$, into the gravitational potential and derive the
modified Friedmann equations via thermodynamic approach at the
apparent horizon of a Friedmann-Robertson-Walker (FRW) universe.
The resulting framework introduces a dimensionless coupling
parameter $\beta\sim l_0^2H_0^2$ quantifying deviations from the
standard $\Lambda$CDM model. Using Bayesian inference with
\textsc{Cobaya} and MCMC sampling, we constrain the model
parameter against late-time observations, including PantheonPlus
and Union3 Type~Ia supernovae, cosmic chronometers, DESI~DR2 BAO
measurements, and Amati-calibrated GRBs. The joint analysis yields
an upper bound $\beta \lesssim \mathcal{O}(10^{-3})$ (68\% C.L.),
implying that departures from $\Lambda$CDM are extremely small
within current precision. Model comparison through the Akaike
Information Criterion shows that the $\Lambda$CDM and T-duality
models provide statistically equivalent fits to the data,
exhibiting only a marginal preference for $\Lambda$CDM. These
results provide the first quantitative observational constraints
on string T-duality inspired modified cosmology and underscore the
potential of future high-precision surveys to test quantum-gravity
induced corrections in a late-time universe.
\end{abstract}
\date{\today}
\maketitle


\section{Introduction}\label{sec1}

In the past few decades, it has become widely accepted that the
laws of gravity may be viewed as a macroscopic manifestation of
the laws of thermodynamics in large-scale spacetime systems. This
idea has been thoroughly explored, and many studies have shown
that gravitational field equations can be derived from the first
law of thermodynamics applied to the boundary of
spacetime~\cite{Jacobson:1995ab,Padmanabhan:2003gd,Padmanabhan:2009vy,Paranjape:2006ca}.
The deep link between gravity and thermodynamics has been
investigated in several contexts. In particular, in the background
of a FRW cosmology, it has been demonstrated that the Friedmann
equations describing the dynamics of the Universe can be written
as the first law of thermodynamics on the apparent horizon, and
vice
versa~\cite{Frolov:2002va,Cai:2005ra,Akbar:2006kj,Cai:2006rs,Sheykhi:2007zp,Sheykhi:2007gi}.

Although General Relativity (GR) has passed many experimental and
observational tests, it still faces major challenges. In
particular, it predicts curvature singularities such as those at
the centers of black holes or at the beginning of the Universe.
Removing these singularities using first principles from
fundamental theories like string theory or quantum gravity remains
one of the key goals of the modern theoretical physics.

Different approaches have been proposed to tackle this problem.
One is the construction of regular black holes, obtained by
coupling nonlinear electrodynamics to the Einstein-Hilbert
action~\cite{Ayon-Beato:1998hmi}. Another method is based on
noncommutative geometry, where spacetime coordinates are averaged
over coherent states inspired by string
theory~\cite{Nicolini:2005vd,Nicolini:2008aj}. In this framework,
noncommutativity leads to a nonlocal modification of the
Einstein-Hilbert action~\cite{Nicolini:2019irw}, which can smooth
out singularities and change the near-horizon structure.

A different and elegant way to obtain nonsingular black holes
arises from T-duality in string theory. In this case, the
momentum-space propagator for a particle of mass $m_0$, modified
by path-integral duality, takes the
form~\cite{Padmanabhan:1996ap,Fontanini:2005ik}
\begin{equation}
\label{G1}
G(p)=-\frac{l_0}{\sqrt{p^2+m_0^{2}}}K_{1}\left(l_0\sqrt{p^2+m_0^{2}}\right),
\end{equation}
where $p^2=(\hbar k)^2$ is the squared momentum, $l_0$ is the
zero-point length of spacetime, and $K_{1}(x)$ is the modified
Bessel function of the second kind. For massless particles
(setting $\hbar=1$), this expression becomes
$G(k)=-\dfrac{l_0}{\sqrt{k^2}}K_{1}\left(l_0\sqrt{k^2}\right)$,
which reduces to the standard propagator $G(k)=-k^{-2}$ in the
limit of vanishing zero-point length.

In compliance with the entropic force picture introduced by
Verlinde~\cite{Verlinde:2010hp}, the inclusion of a fundamental
length scale in the gravitational interaction was first examined
in~\cite{Jusufi:2022mir}, where corrections to Newton's law of
gravity were derived. These modifications give rise to departures
from the classical inverse-square dependence, effectively
regularizing the gravitational field at short distances. As a
result, the potential remains finite at the origin, generating a
regular geometry without curvature
singularities~\cite{Nicolini:2019irw}. Such a mechanism also
impacts black hole thermodynamics, yielding modified expressions
for the Hawking temperature.

Nevertheless, given that cosmology probes physical processes at
energy scales far beyond those accessible in local gravitational
systems, it appears natural to investigate phenomenological
consequences of the zero-point length within this framework. On
cosmological scales, the combination of the modified gravitational
potential and Verlinde's conjecture yields correction terms to the
Newtonian formulation of cosmology and, in the relativistic
regime, leads to modified Friedmann equations in a FRW
background~\cite{Jusufi:2022mir}. This approach also results in a
modified expression for the entropy associated with the
cosmological horizon, arising from the introduction of the
zero-point length. It is worth noting that the assumption of a
modified horizon entropy is a common feature among various
extensions of general
relativity~\cite{Tsallis:2013,Barrow:2020tzx,Nojiri:2019skr,Lymperis:2018iuz,Saridakis:2020lrg,Hernandez-Almada:2021rjs,Dheepika:2022sio,Jizba:2022icu,Lambiase:2023ryq,Jizba:2024klq,Ebrahimi:2024zrk,Nojiri:2025gkq,Luciano:2022pzg,Ghaffari:2022skp,Basilakos:2025wwu},
and similar considerations have also appeared in particle physics
contexts~\cite{Kaniadakis:2002zz,Kapusta:2021zfo,
Luciano:2021mto}.

Inspired by string T-duality and incorporating the zero-point
length correction \(l_0\) into the gravitational potential,
Ref.~\cite{Luciano:2024mcn} formulated modified Friedmann
equations by applying the first law of thermodynamics to the
apparent horizon of the FRW Universe. The analysis revealed that
\(l_0\) is constrained around the Planck scale and affects the
early-Universe dynamics in a non-trivial manner. Moreover, the
framework provides an alternative interpretation for the
broken-power-law spectrum and suggests that imprints of the
zero-point length could manifest in the primordial
gravitational-wave background, offering a potential test of
deviations from GR in future experiments.

On the observational side, the recent data releases from the Dark
Energy Spectroscopic Instrument (DESI), including its upgraded
DESI DR2, have consolidated their role as one of the most precise
cosmological probes currently available
\cite{DESI:2025zpo,DESI:2025zgx,DESI:2025fii}. The high-quality
BAO measurements extracted from DESI data provide robust
constraints on the expansion history of the Universe and are
routinely employed across a broad spectrum of theoretical
frameworks, ranging from standard and dynamical dark energy
models~\cite{Ormondroyd:2025iaf,You:2025uon,Gu:2025xie,Santos:2025wiv,Li:2025cxn,Carloni:2024zpl,Luciano:2025elo,Luciano:2025hjn}
to alternative scenarios such as early dark
energy~\cite{Chaussidon:2025npr}, scalar-tensor theories with
minimal or non-minimal
couplings~\cite{Anchordoqui:2025fgz,Ye:2025ulq,Paliathanasis:2025xxm,Arora:2025msq},
and extensions inspired by the Generalized Uncertainty
Principle~\cite{Paliathanasis:2025dcr}. They also play a key role
in testing interacting dark sector
models~\cite{Shah:2025ayl,Silva:2025hxw,vanderWesthuizen:2025iam,Guedezounme:2025wav,
Li:2024qso}, constraining astrophysical
relations~\cite{Alfano:2024jqn}, and performing cosmographic
analyses~\cite{Luongo:2024fww}, as well as in a variety of
modified gravity
contexts~\cite{Yang:2025mws,Paliathanasis:2025hjw,Tyagi:2025zov,Li:2025dwz,Hogas:2025ahb}.

In parallel, complementary datasets such as the PantheonPlus
(PP)~\cite{Scolnic:2021amr} and Union3 (U3)~\cite{Rubin:2023jdq}
Type Ia supernova (SNIa) compilations, the cosmic chronometer (CC)
determinations of the Hubble (OHD)
parameter~\cite{Vagnozzi:2020dfn}, and the Amati-calibrated
gamma-ray burst (GRB) sample~\cite{Amati:2002ny, Amati:2018tso}
offer independent and model-agnostic measurements of cosmic
distances over a wide redshift range. When analyzed jointly with
DESI data, these probes provide a consistent and highly reliable
picture of the late-time cosmic expansion, enabling a detailed
comparison among diverse cosmological scenarios. Furthermore, this
synergy among BAO, SNI and GRB datasets has become a standard
strategy in current precision cosmology, allowing for stringent
tests of both conventional and quantum-gravity–motivated models.

Starting from these premises, in this work we aim to extend the
investigation of string T-duality from a phenomenological and
observational perspective. Specifically, we confront the
theoretical predictions with the latest cosmological datasets,
namely DESI DR2, PP, U3, OHD and GRB. The combination of these
probes enables a robust statistical analysis and a direct
comparison with the standard \(\Lambda\)CDM cosmology, thereby
assessing the viability of the T-duality model as a
phenomenological framework potentially sensitive to
quantum-gravity effects at cosmological scales.

The paper is organized as follows. In Sec.~\ref{sec2}, we outline
the theoretical foundations of the extended cosmology emerging
from string T-duality, following the treatment presented in
Ref.~\cite{Luciano:2024mcn}. Section~\ref{sec3} is dedicated to
the analysis of observational data, whereas the main conclusions
and future perspectives are discussed in Sec.~\ref{sec4}. Unless
stated otherwise, natural units are used throughout this work.
\section{Cosmology from string T-duality}
\label{sec2} According to the T-duality principle, the existence
of a zero-point length \(l_0\) induces a modification in the
gravitational potential, effectively regularizing the interaction
at short distances. Within the entropic gravity framework, this
deformation translates into a correction to the entropy associated
with the cosmological horizon. The resulting differential form of
the horizon entropy can be expressed as~\cite{Luciano:2024mcn}
\begin{equation}
\label{dS}
dS_h = 2\pi R \left(1 + \frac{l_0^2}{R^2}\right)^{-3/2} dR\,.
\end{equation}
Here, the radial coordinate \(R\) admits a twofold interpretation.
In static, spherically symmetric configurations, it represents the
distance from the center of a mass distribution \(M\), at which
the modified potential and entropic force are evaluated. In the
cosmological setting, \(R\) is instead identified with the radius
of the apparent horizon of the FRW Universe, thus linking the
local modification of gravity induced by the zero-point length to
the global thermodynamic behavior of the expanding spacetime. In
the limit \(l_0/R \to 0\), Eq.~\eqref{dS} naturally reduces to the
standard Bekenstein-Hawking entropy relation, $S_{BH}=\pi R^2$.

Let us assume that the Universe is described by the FRW line element
\begin{equation}
ds^2 = h_{\mu\nu}dx^{\mu}dx^{\nu} + R^2(d\theta^2 + \sin^2\theta\, d\phi^2),
\end{equation}
where \(R = a(t)r\), \(x^0 = t\), \(x^1 = r\), and \(h_{\mu\nu} =
\mathrm{diag}(-1,\, a^2/(1 - k r^2))\) denotes the two-dimensional
metric. The parameter \(k = -1, 0, +1\) corresponds to open, flat,
and closed spatial geometries, respectively, while \(a(t)\)
represents the scale factor.

From a thermodynamic standpoint, the apparent horizon serves as
the appropriate causal boundary of the Universe, with radius
\begin{equation}
\label{radius}
R = \frac{1}{\sqrt{H^2 + k/a^2}}.
\end{equation}
For computational simplicity, and in light of recent observational
constraints that strongly support a spatially flat Universe, we
shall henceforth consider the case of vanishing curvature,
$k=0$~\cite{Planck:2018vyg}.

In turn, the temperature associated with this horizon is given by~\cite{Akbar:2006kj}
\begin{equation}\label{Th}
T_h = -\frac{1}{2\pi R}\!\left(1 - \frac{\dot{R}}{2HR}\right),
\end{equation}
where a dot denotes differentiation with respect to cosmic time \(t\).

The total energy-momentum tensor is modeled as a perfect fluid,
$T_{\mu\nu} = (\rho + p)u_{\mu}u_{\nu} + p\,g_{\mu\nu}$, where
\(\rho\) and \(p\) denote the energy density and pressure,
respectively. Conservation of energy-momentum then yields the
continuity equation
\begin{equation}
\label{Cont}
\dot{\rho} + 3H(\rho + p) = 0,
\end{equation}
with \(H = \dot{a}/a\) being the Hubble parameter.
The work density associated with the cosmic volume variation is defined as~\cite{Hayward:1997jp}
\begin{equation}
\label{Work2}
W = -\tfrac{1}{2}T^{\mu\nu}h_{\mu\nu} = \frac{1}{2}(\rho - p).
\end{equation}
Considering the total energy \(E = \rho V = \rho (4\pi R^3 / 3)\) and applying the conservation law~\eqref{Cont}, one obtains
\begin{equation}
\label{dE2}
dE = 4\pi R^2\rho\, dR - 4\pi H R^3 (\rho + p)\,dt.
\end{equation}

By inserting Eqs.~\eqref{dS}, \eqref{Th}, \eqref{Work2} and
\eqref{dE2} into the first law of thermodynamics at the apparent
horizon, $dE = T_h dS_h + W dV$, one arrives, after
straightforward algebra, at the modified Friedmann
equation~\cite{Luciano:2024mcn}
\begin{equation}
\label{Frie6}
H^2 - \alpha H^4 + \mathcal{O}(\alpha^2)
= \frac{8\pi}{3}\rho + \frac{\Lambda}{3},
\end{equation}
where $\Lambda$ denotes the cosmological constant and $\alpha
\equiv 3l_0^2/4$. Thus, the introduction of the modified
entropy~\eqref{dS} naturally leads to a correction in the
Friedmann dynamics, producing an additional \(H^4\) term beyond
the standard GR result. The latter is consistently recovered for
$\alpha\rightarrow0$.

\subsection{Cosmological solutions}
We now turn to the cosmological implications of the zero-point
length framework. In what follows, we consider a Universe filled
with a perfect fluid composed of dust matter (\(p_m \simeq 0\))
and radiation. From the continuity equation, these components
evolve as $\rho_m = \rho_{m,0}\, a^{-3}$ and $\rho_r =
\rho_{r,0}\, a^{-4}$, respectively, where the subscript “0’’
denotes present-day values, and the scale factor is normalized to
\(a_0 = 1\).

Dividing both sides of Eq.~\eqref{Frie6} by \(H_0^2\), one obtains
\begin{equation}
\label{Frim1}
-\beta\!\left(\frac{H}{H_0}\right)^4
+ \left(\frac{H}{H_0}\right)^2
- \Omega_{m,0} a^{-3}
- \Omega_{r,0} a^{-4}
- \Omega_{\Lambda,0} = 0\,,
\end{equation}
where the dimensionless coupling parameter is defined as
\begin{equation}
\label{beta}
\beta \equiv \alpha H_0^2 = \frac{3}{4}\, l_0^2 H_0^2 \ll 1\,,
\end{equation}
and the standard density parameters are introduced via
\begin{equation}
\Omega_i = \frac{8\pi}{3H^2}\rho_i\,, \quad (i = m, r)\,, \qquad\,\,\, \Omega_{\Lambda} = \frac{\Lambda}{3H^2}\,.
\end{equation}

Solving Eq.~\eqref{Frim1} for \(H/H_0\) yields
\begin{equation}
\label{Frim2}
\left[\frac{H(z)}{H_0}\right]^2 =
\frac{1}{2\beta}
\left\{
1 - \sqrt{1 - 4\beta
\left[\Omega_{m,0}\left(1+z\right)^3 + \Omega_{r,0}\left(1+z\right)^4 + \Omega_{\Lambda,0}\right]}
\right\}\,,
\end{equation}
where we have retained the only solution that yields the correct behavior in the limit $\beta \rightarrow 0$.
Furthermore, we have introduced the redshift parameter through $a^{-1}=\left(1+z\right)$.

Since \(\beta\) is expected to be small, the right-hand side can be expanded to first order, leading to
\begin{equation}
\label{Frim6}
\left[\frac{H(z)}{H_0}\right]^2 =
\left[
\Omega_{m,0} (1+z)^3 + \Omega_{r,0} (1+z)^4 + \Omega_{\Lambda,0}
\right] \left\{
1 + \beta\!\left[
\Omega_{m,0} (1+z)^3 + \Omega_{r,0} (1+z)^4 + \Omega_{\Lambda,0}
\right]
\right\} + \mathcal{O}(\beta^2)\,.
\end{equation}
Finally, imposing the flatness condition \(H(0)/H_0 = 1\) determines the present-day density parameter as
\begin{equation}
\label{flatness}
\Omega_{\Lambda,0} =
1 - \Omega_{m,0} - \Omega_{r,0} - \beta + \mathcal{O}(\beta^2)\,.
\end{equation}
\section{Observational Data Analysis}\label{sec3}
In this section, we present the observational constraints obtained
for our string T-duality cosmological model and compare the
resulting physical parameters with those of the standard
$\Lambda$CDM model. We also describe the cosmological datasets
employed in the analysis and outline the statistical methods used
for parameter estimation and model comparison.
\subsection{Observational Data}
In the following sections, we present the observational datasets
utilized in this study.
\begin{itemize}
\item Supernova PantheonPlus (PP): We consider the PP catalogue for supernova
data~\cite{pan}. The catalogue includes 1701 light curves of 1550
spectroscopically confirmed events within the range $10^{-3} < z < 2.27$. The data
provide observable values for the distance modulus $\mu^{\mathrm{obs}}$. The
theoretical distance modulus~$\mu^{\mathrm{th}}$ is constructed from the luminosity
distance $D_{L}(z)$, which, for a spatially flat FLRW geometry, is defined as
\begin{equation}
D_{L}(z) = (1+z)\int_{0}^{z}\frac{dz^{\prime}}{H(z^{\prime})}\,.
\end{equation}
We employ
the PantheonPlus catalogue without the SH0ES Cepheid calibration.

\item Supernova Union3 (U3): We consider a second set of supernova data, provided by the U3 catalogue~\cite{union}. It includes 2087 supernova events
within the same redshift range as the PP data. A total of 1363 events are common with
the PP catalogue.

\item Observational Hubble Data (OHD): We apply the direct measurements of the
Hubble parameter obtained from Cosmic Chronometers (CC). These measurements
are model-independent, as they do not rely on any specific cosmological
assumptions. CC are passively evolving galaxies with synchronous stellar
populations and comparable cosmic evolution~\cite{co01}. In this work, we use
31 direct measurements of the Hubble parameter over the redshift range
$0.09 \leq z \leq 1.965$, as presented in~\cite{cc1}.

\item Baryonic Acoustic Oscillations (BAO): We make use of the latest baryon
acoustic oscillation (BAO) data from the Dark Energy Spectroscopic Instrument
(DESI DR2)~\cite{DESI:2025zpo,DESI:2025zgx,DESI:2025fii}. This dataset provides measurements of the
transverse comoving angular distance ratio,
\begin{equation}
\frac{D_{M}}{r_{drag}} =
\frac{D_{L}}{(1+z)\, r_{drag}}\,,
\end{equation}
the volume-averaged distance
ratio,
\begin{equation}
\frac{D_{V}}{r_{drag}} =
\frac{(z D_{H} D_{M}^{2})^{1/3}}{r_{drag}}\,,
\end{equation}
and the Hubble distance ratio,
\begin{equation}
\frac{D_{H}}{r_{drag}} =
\frac{1}{r_{drag} H(z)}\,,
\end{equation}
at seven distinct redshifts, where $D_{L}$
refers to the luminosity distance and $r_{drag}$ denotes the sound horizon at
the drag epoch. In this study, $r_{drag}$ is treated as a free parameter.

\item Gamma-ray Bursts (GRBs): We consider 193 events analyzed using the Amati correlation~\cite{amm} within the redshift range $0.0335 < z < 8.1$.
The Amati correlation is a model-independent approach that allows GRBs to be used as distance indicators. The distance modulus $\mu^{\mathrm{obs}}$
for each measurement at the observed redshift, after calibration, is presented in~\cite{Amati:2018tso}.
\end{itemize}
\subsection{Methodology}
To perform the statistical analysis, we adopt the Bayesian
inference framework provided by
COBAYA\footnote{\url{https://cobaya.readthedocs.io/}}~\cite{cob1,cob2},
using a custom theory implementation together with the MCMC
sampler~\cite{Lewis:2002ah,Lewis:2013hha}. The MCMC chains are
analyzed with the GetDist
library\footnote{\url{https://getdist.readthedocs.io/}}~\cite{Lewis:2019xzd}.

We consider six different combinations of datasets, as summarized
in Table~\ref{bestfit}. For each dataset, we determine the
best-fit parameters that maximize the likelihood,
$\mathcal{L}_{\max} = \exp\left(-\frac{1}{2}
\chi_{\min}^{2}\right)$, where
\begin{equation}
\chi_{\min}^{2} = \chi_{\min(\mathrm{data1})}^{2} + \chi_{\min(\mathrm{data2})}^{2} + \ldots~.
\end{equation}
In order to perform a statistical comparison between the string
T-duality cosmology and the $\Lambda$CDM model, we employ the
Akaike Information Criterion (AIC)~\cite{AIC}. Akaike's scale
allows us to compare models with different degrees of freedom by
using the AIC. The latter, for a large number of data points, is
defined in terms of the minimum chi-squared value,
$\chi_{\min}^{2}$, and the number of free parameters, $\kappa$, as
follows:
\begin{equation}
\mathrm{AIC} \simeq \chi_{\min}^{2} + 2\kappa.
\end{equation}
The free parameters for the $\Lambda$CDM model are $\{ H_{0},
\Omega_{m0}, r_{drag} \}$, while for the string T-duality
cosmology they are $\{ H_{0}, \Omega_{m0}, r_{drag}, \beta \}$. In
Table~\ref{prior}, we present the priors considered for the MCMC
sampler.

Akaike's scale uses the difference in the AIC values of the two models, i.e.,
$\Delta \mathrm{AIC} = \mathrm{AIC}_{\text{model1}} - \mathrm{AIC}_{\text{model2}}$.
For $\lvert \Delta \mathrm{AIC} \rvert < 2$, the two models are statistically equivalent.
However, for $2 < \lvert \Delta \mathrm{AIC} \rvert < 6$, there is weak evidence in favor
of the model with the smaller AIC value; if $6 < \lvert \Delta \mathrm{AIC} \rvert < 10$,
the evidence is strong; and for $\lvert \Delta \mathrm{AIC} \rvert > 10$, there is clear
evidence supporting a preference for the model with the lower AIC.
\begin{table}[t]
\centering
\caption{Priors of the Free Parameters.}%
\begin{tabular}
[c]{ccc}\hline\hline
\textbf{Priors} & \textbf{T-Duality} & $\Lambda$\textbf{CDM}\\\hline
$\mathbf{H}_{0}\, (\mathrm{km\, s^{-1} Mpc^{-1}})$ & $\left[  60,80\right]  $ & $\left[  60,80\right]  $\\
$\mathbf{\Omega}_{m0}$ & $\left[  0.01,0.5\right]  $ & $\left[
0.01,0.5\right]  $\\
$\mathbf{\beta}$ & $\left[  0,1\right]  $ & $-$\\
$\mathbf{r}_{drag}\, (\mathrm{Mpc})$ & $\left[  130,160\right]  $ & $\left[  130,160\right]
$\\\hline\hline
\end{tabular}
\label{prior}%
\end{table}%

\subsection{Results}
We perform six different constraints corresponding to the dataset
combinations summarized in Table~\ref{bestfit}. The best-fit
parameters, together with the statistical comparison to the
$\Lambda$CDM model, are presented in Table~\ref{bestfit}. In what
follows, we discuss the results for each combination, with
particular emphasis on the T-duality coupling parameter~$\beta$.

For the dataset $\mathbf{D}_{1}$, which combines the PP and OHD data,
we obtain $H_{0} = 67.4^{+1.8}_{-1.8}$ and $\Omega_{m0} = 0.32^{+0.019}_{-0.019}$,
with the upper bound $\beta < 0.0122$. The comparison with the $\Lambda$CDM
model gives $\chi_{\min}^{2} - \chi_{\Lambda,\min}^{2} = +0.2$,
corresponding to $\Delta \mathrm{AIC} = +2.2$. According to Akaike’s
scale, this indicates weak evidence in favor of the $\Lambda$CDM model,
though both models remain statistically consistent.

Including BAO measurements in dataset $\mathbf{D}_{2}$
(PP + OHD + BAO) improves the constraints, yielding
$H_{0} = 68.7^{+1.6}_{-1.6}$, $\Omega_{m0} = 0.30^{+0.013}_{-0.013}$,
$r_{drag} = 146.9^{+3.0}_{-3.5}$ and
$\beta < 0.0030$. The statistical comparison gives
$\chi_{\min}^{2} - \chi_{\Lambda,\min}^{2} = 0.0$ and
$\Delta \mathrm{AIC} = +2.0$, indicating that the model fits the data equally well as the $\Lambda$CDM model.

For dataset $\mathbf{D}_{3}$, which further includes GRB data (PP
+ OHD + BAO + GRB), the best-fit parameters are $H_{0} =
69.3^{+1.7}_{-1.7}$, $\Omega_{m0} = 0.29^{+0.013}_{-0.012}$,
$r_{drag} = 147.0^{+3.4}_{-3.4}$, and $\beta < 0.0034$. The
inclusion of GRBs does not improve the constraint on~$\beta$, with
$\chi_{\min}^{2} - \chi_{\Lambda,\min}^{2} = 0$ and $\Delta
\mathrm{AIC} = +2.0$. Therefore, the models are statistically
equivalent.

Replacing the PP catalogue with the U3 supernova sample allows us to test the robustness of our results using an updated and extended dataset.
For the dataset $\mathbf{D}{4}$ (U3 + OHD), we obtain
$H_{0} = 66.8^{+1.9}_{-1.9}$, $\Omega_{m0} = 0.34^{+0.025}_{-0.025}$ and $\beta < 0.0111$.
The comparison yields $\chi_{\min}^{2} - \chi_{\Lambda,\min}^{2} = 0.0$ and $\Delta \mathrm{AIC} = +2.0$.
Therefore, the data do not favor either model over the other.

For dataset $\mathbf{D}_{5}$ (U3 + OHD + BAO), the constraints
tighten considerably, with
$H_{0} = 68.7^{+1.7}_{-1.7}$, $\Omega_{m0} = 0.30^{+0.015}_{-0.015}$,
$r_{drag} = 146.9^{+3.4}_{-3.4}$, and $\beta < 0.0033$.
The resulting $\chi_{\min}^{2} - \chi_{\Lambda,\min}^{2} = +0.1$ and
$\Delta \mathrm{AIC} = +2.1$ confirm a weak preference for the
$\Lambda$CDM model.

Finally, for dataset $\mathbf{D}_{6}$ (U3 + OHD + BAO + GRB), we
obtain $H_{0} = 69.4^{+1.6}_{-1.6}$, $\Omega_{m0} =
0.29^{+0.015}_{-0.013}$, $r_{drag} = 147.3^{+3.3}_{-3.3}$, and
$\beta < 0.0042$. In this case, the model comparison yields
$\chi_{\min}^{2} - \chi_{\Lambda,\min}^{2} = 0.0$ and $\Delta
\mathrm{AIC} = +2.0$, indicating no significant statistical
preference between the models.

\begin{figure}[t]
\centering\includegraphics[width=0.7\textwidth]{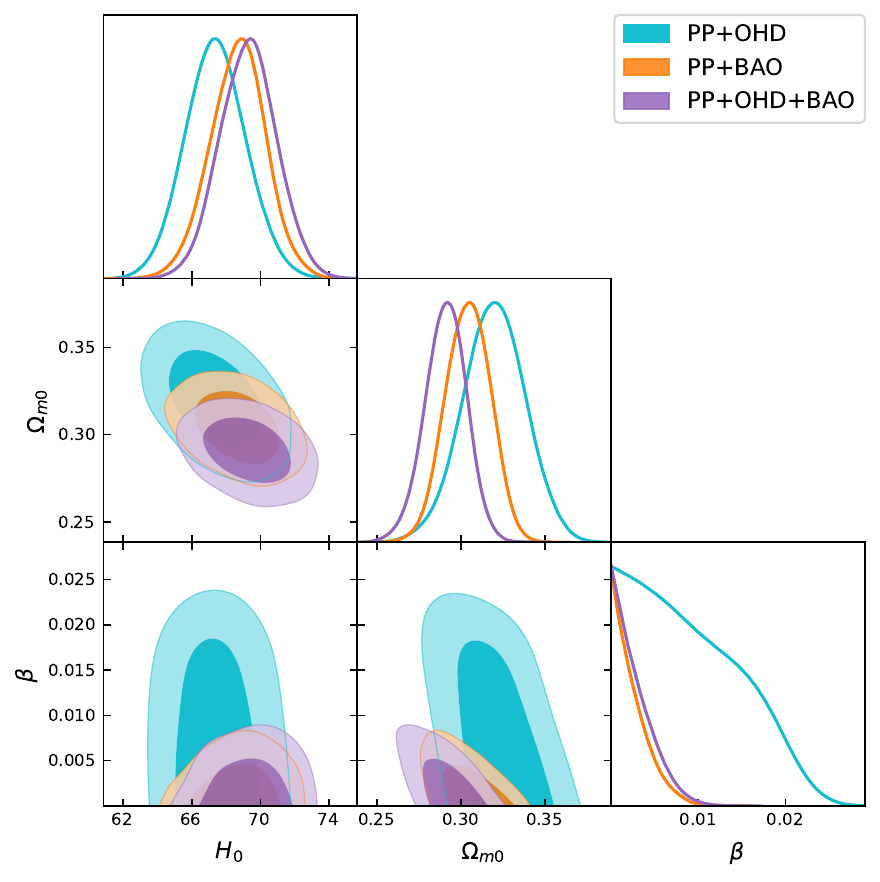}\caption{Confidence
regions for the free parameters of the string T-duality cosmology, derived
from the joint likelihood of the combined datasets including the PP SNIa
catalogue. }%
\label{fig1}%
\end{figure}

\begin{figure}[t]
\centering\includegraphics[width=0.7\textwidth]{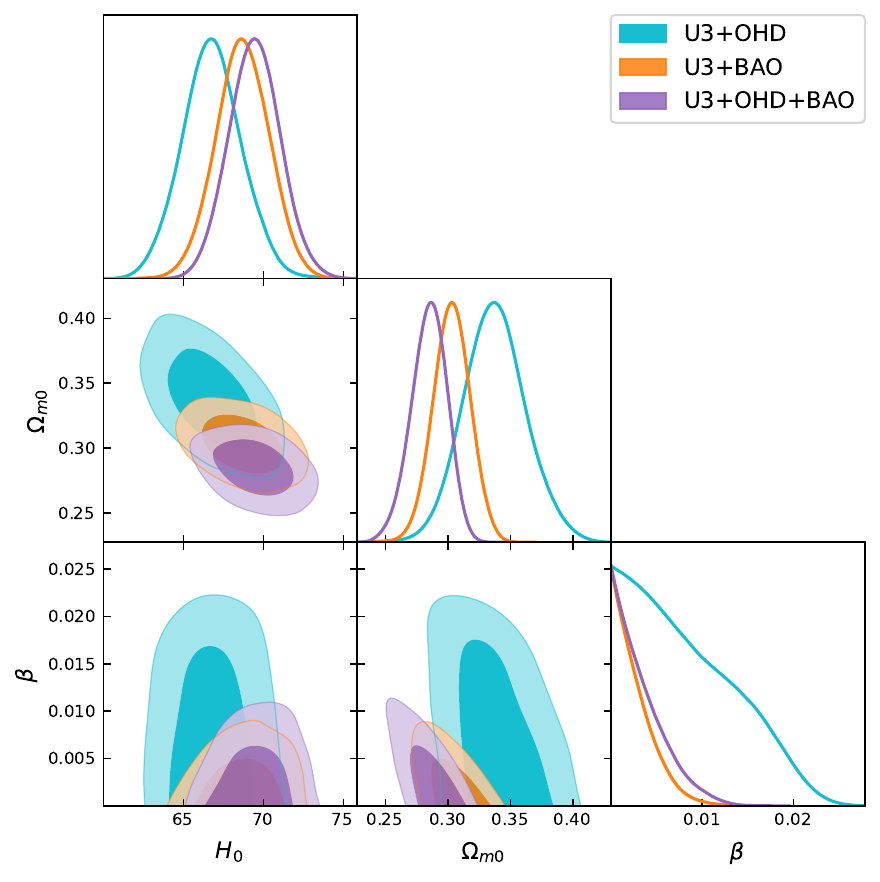}\caption{Confidence
regions for the free parameters of the string T-duality cosmology, derived
from the joint likelihood of the combined datasets including the U3 SNIa
catalogue. }%
\label{fig2}%
\end{figure}

In Figs.~\ref{fig1} and \ref{fig2} we show the contour plots of
the confidence regions for the best-fit parameters of the string
T-duality model. In summary, the inclusion of BAO data plays a
decisive role in tightening the upper bounds on the T-duality. The
addition of GRB measurements, on the other hand, does not lead to
significant improvements. Overall, the AIC analysis suggests that
the $\Lambda$CDM model remains slightly favored, while the small
upper limits on~$\beta$ indicate that any deviations from standard
cosmology are minimal.




\begin{table}[tbp] \centering
\caption{Observational Constraints for the string T-duality cosmological model and the $\Lambda$CDM.}%
\begin{tabular}
[c]{ccccccc}\hline\hline
& $\mathbf{H}_{0}$ & $\mathbf{\Omega}_{m0}$ & $\mathbf{r}_{drag}$ &
$\mathbf{\beta}$ & $\mathbf{\chi}_{\min}^{2}\mathbf{-\chi}_{\Lambda,\min}^{2}$
& $\mathbf{AIC-AIC}_{\Lambda}$\\\hline
\multicolumn{7}{c}{\textbf{Dataset: }$\mathbf{PP\&OHD}$}\\
\textbf{T-Duality} & $67.4_{-1.8}^{+1.8}$ & $0.32_{-0.019}^{+0.019}$ & $-$ &
$<0.0122$ & $+0.2$ & $+2.2$\\
$\Lambda$\textbf{CDM} & $67.8_{-1.8}^{+1.8}$ & $0.33_{-0.018}^{+0.018}$ & $-$
& $-$ & $0$ & $0$\\
\multicolumn{7}{c}{\textbf{Dataset}: $\mathbf{PP\&OHD\&BAO}$}\\
\textbf{T-Duality} & $68.7_{-1.6}^{+1.6}$ & $0.30_{-0.013}^{+0.013}$ &
$146.9_{-3.5}^{+3.0}$ & $<0.0030$ & $+0.0$ & $+2.0$\\
$\Lambda$\textbf{CDM} & $68.5_{-1.6}^{+1.6}$ & $0.31_{-0.012}^{+0.011}$ &
$147.1_{-3.4}^{+3.4}$ & $-$ & $0$ & $0$\\
\multicolumn{7}{c}{\textbf{Dataset}: $\mathbf{PP\&OHD\&BAO\&GRBs}$}\\
\textbf{T-Duality} & $69.3_{-1.7}^{+1.7}$ & $0.29_{-0.012}^{+0.013}$ &
$147.0_{-3.4}^{+3.4}$ & $<0.0034$ & $+0.0$ & $+2.0$\\
$\Lambda$\textbf{CDM} & $69.0_{-1.7}^{+1.7}$ & $0.30_{-0.011}^{+0.011}$ &
$147.1_{-3.5}^{+3.5}$ & $-$ & $0$ & $0$\\
\multicolumn{7}{c}{\textbf{Dataset: }$\mathbf{U3\&OHD}$}\\
\textbf{T-Duality} & $66.8_{-1.9}^{+1.9}$ & $0.34_{-0.025}^{+0.025}$ & $-$ &
$<0.0111$ & $+0.0$ & $+2.0$\\
$\Lambda$\textbf{CDM} & $66.8_{-1.9}^{+1.9}$ & $0.35_{-0.025}^{+0.025}$ & $-$
& $-$ & $0$ & $0$\\
\multicolumn{7}{c}{\textbf{Dataset}: $\mathbf{U3\&OHD\&BAO}$}\\
\textbf{T-Duality} & $68.7_{-1.7}^{+1.7}$ & $0.30_{-0.015}^{+0.015}$ &
$146.9_{-3.4}^{+3.4}$ & $<0.0033$ & $+0.1$ & $+2.1$\\
$\Lambda$\textbf{CDM} & $68.6_{-1.6}^{+1.6}$ & $0.31_{-0.014}^{+0.014}$ &
$146.7_{-3.4}^{+3.4}$ & $-$ & $0$ & $0$\\
\multicolumn{7}{c}{\textbf{Dataset}: $\mathbf{U3\&OHD\&BAO\&GRBs}$}\\
\textbf{T-Duality} & $69.4_{-1.6}^{+1.6}$ & $0.29_{-0.013}^{+0.015}$ &
$147.3_{-3.3}^{+3.3}$ & $<0.0042$ & $0$ & $+2.0$\\
$\Lambda$\textbf{CDM} & $69.3_{-1.8}^{+1.8}$ & $0.30_{-0.012}^{+0.012}$ &
$146.9_{-3.6}^{+3.6}$ & $-$ & $0$ & $0$\\\hline\hline
\end{tabular}
\label{bestfit}%
\end{table}%

\section{Discussion and Conclusions}\label{sec4}
In this work, we have investigated the phenomenological
implications of the string T-duality framework in late-time
cosmology. By incorporating the zero-point length corrections
motivated by T-duality into the Friedmann equations, we have
derived a modified cosmological model characterized by the
dimensionless coupling parameter~$\beta$, which quantifies
deviations from the standard $\Lambda$CDM dynamics.

Using a comprehensive set of recent cosmological observations -
including the PantheonPlus (PP) and Union3 (U3) Type~Ia supernova
samples, cosmic chronometer (OHD) measurements, baryon acoustic
oscillations (BAO) from DESI DR2, and calibrated gamma-ray bursts
(GRBs) - we performed a joint Bayesian analysis based on MCMC
sampling within the \texttt{Cobaya} framework. Six different
combinations of datasets were explored to assess the robustness of
the constraints and to evaluate potential degeneracies among the
cosmological parameters.

Our results show that the string T-duality cosmology provides an
excellent fit to the observational data, remaining statistically
consistent with the $\Lambda$CDM model across all dataset
combinations. The coupling parameter~$\beta$ is tightly
constrained, with upper limits at the level of $\beta \lesssim
\mathcal{O}(10^{-2})$ from supernova and Hubble rate measurements,
and stronger bounds of $\beta \lesssim \mathcal{O}(10^{-3})$ once
BAO data are included. The addition of GRB data does not yield
substantial improvements, reflecting the current limitations in
GRB statistics and calibration precision.

Although late-time cosmological observations provide bounds on the
minimal length  that are less stringent than those derived in
high-energy contexts such as quantum-corrected black holes
\cite{Nicolini:2019irw} or the primordial gravitational-wave
spectrum \cite{Luciano:2024mcn}, they may nevertheless play a
crucial and complementary role. Indeed, these observations probe
the expansion history of the Universe with high precision and
offer an independent, low-energy consistency test of the T-duality
framework within the cosmological regime.

Future investigations may extend this analysis toward early-time
cosmology, where T-duality effects are expected to leave more
pronounced imprints. In particular, the cosmic microwave
background (CMB) provides a sensitive probe of high-energy
modifications through their influence on the acoustic peaks,
lensing potential and polarization spectra. A combined analysis of
these early- and late-time datasets would therefore enable a more
comprehensive test of the T-duality paradigm, bridging the gap
between quantum-gravity-inspired theories and precision cosmology.

In parallel, an intriguing avenue for future work is to
reformulate the present analysis within the framework of
alternative minimal-length approaches, in order to clarify
possible similarities and differences. Promising candidates
include the q-metric
formalism~\cite{Kothawala:2013maa,Kothawala:2014tya,Pesci:2018syy},
which naturally embeds a minimal length in the spacetime geometry
through a generalized definition of metric distances, and the
Generalized Uncertainty Principle (GUP), which introduces a
Planck-scale modification of the Heisenberg uncertainty relations
due to quantum-gravity effects \cite{Kempf:1994su,Bosso:2023aht}.
A comparative study among these frameworks could help identify
universal phenomenological features associated with the existence
of a minimal length, and provide valuable guidance toward a
unified understanding of quantum-spacetime effects across
different physical regimes. Work along these directions is in
progress and will be explored elsewhere.
\begin{acknowledgments}
The research of GGL is supported by the postdoctoral fellowship program of the
University of Lleida. GGL gratefully acknowledges the contribution of the LISA
Cosmology Working Group (CosWG), as well as support from the COST Actions
CA21136 - \textit{Addressing observational tensions in cosmology with
systematics and fundamental physics (CosmoVerse)} - CA23130, \textit{Bridging
high and low energies in search of quantum gravity (BridgeQG)} and CA21106 -
\textit{COSMIC WISPers in the Dark Universe: Theory, astrophysics and
experiments (CosmicWISPers)}. AP thanks the support of VRIDT through
Resoluci\'{o}n VRIDT No. 096/2022 and Resoluci\'{o}n VRIDT No. 098/2022. Part
of this study was supported by FONDECYT 1240514.
\end{acknowledgments}

\bibliography{Bib}

\end{document}